# Permit Allocation in Emissions Trading Using the Boltzmann Distribution

Ji-Won Park [a], Chae Un Kim [b], and Walter Isard [ac]

[a] Regional Science, [b] Cornell High Energy Synchrotron Source (CHESS), [c] Economics, Cornell University, Ithaca, NY 14853, USA.

**Author names and affiliations**

Ji-Won Park
106 W. Sibley Hall, Cornell University, Ithaca, NY 14853, U.S.A.
Tel: +1-607-229-2639
Fax: +1-607-254-2848
E-mail: jp429@cornell.edu

Chae Un Kim
200 L Wilson Lab, Cornell University, Ithaca, NY 14853, U.S.A.
Tel: +1-607-255-7163
Fax: +1-607-255-8751
E-mail: ck243@cornell.edu

Walter Isard*
476 Uris Hall, Cornell University, Ithaca, NY 14853, U.S.A.
Tel: +1-607-255-3306
E-mail: wi1@cornell.edu
* Professor Isard passed away before the submission of this manuscript at the age of 91.

**Corresponding author**

Ji-Won Park
106 W. Sibley Hall, Cornell University, Ithaca, NY 14853, U.S.A.
Tel: +1-607-229-2639
Fax: +1-607-254-2848
E-mail: jp429@cornell.edu

**Abstract**

**In emissions trading, the initial allocation of permits is an intractable issue because it needs to be essentially fair to the participating countries. There are many ways to distribute a given total amount of emissions permits among countries, but the existing distribution methods, such as auctioning and grandfathering, have been debated. In this paper we describe a new method for allocating permits in emissions trading using the Boltzmann distribution. We introduce the Boltzmann distribution to permit allocation by combining it with concepts in emissions trading. We then demonstrate through empirical data analysis how emissions permits can be allocated in practice among participating countries. The new allocation method using the Boltzmann distribution describes the most probable, natural, and unbiased distribution of emissions permits among multiple countries. Simple and versatile, this new method holds potential for many economic and environmental applications.**

# 1. Introduction

Scientists have warned that global warming of more than 1°C would constitute a dangerous climate change based on the likely effects on sea levels and the extermination of species [1]. Furthermore, climate change that occurs as a result of increases in $CO_2$ concentration is largely irreversible for 1,000 years, even after $CO_2$ emissions cease [2]. According to the *Stern Review,* prompt, decisive action is clearly warranted, and, because climate change is a global problem, the response to it must be international [3].

Various ideas have been proposed for slowing global warming and reducing $CO_2$ emissions to the atmosphere, including reflecting solar radiation with small particles in the stratosphere, putting deflectors in space, growing trees and other biomass to remove $CO_2$ from the atmosphere, fertilizing oceans with iron to remove $CO_2$, and reducing $CO_2$ emissions through carbon taxes or emissions trading [4]. Among these, numerous studies found that emissions trading lowers the cost of reaching the commitments of the Kyoto Protocol [5].

The basic concepts of emissions trading were established in past decades [6-13]. As with any trading system, in the emissions trading system, the flow and value of what is traded depends on its initial allocation, its supply, and the demand for it [14]. There are many possible ways to distribute a given total of emissions permits among participants [15]; traditionally, grandfathering and auctioning have been suggested for initial permit allocation [16, 17].

Permit allocation is one of the most intractable issues to resolve in designing emissions trading systems. A permit-allocation rule should be simple, should be based in part on historical data, and should be perceived as fair [18]. Because the flow and value of emissions permits depends on their initial allocation, the fair allocation of a limited number of permits among countries or firms is not only important but also controversial.

In this paper, we introduce an alternative method for initial permit allocation using the Boltzmann distribution. We first describe the basic concept of the Boltzmann distribution and then develop its mathematical formula for the allocation of emissions permits. Next, through empirical data analysis, we demonstrate how this allocation method can be used in practice for initial permit allocation.

## 2. The Boltzmann distribution

In the physical sciences, the Boltzmann distribution yields the equilibrium probability distribution of a physical system in its energy substates [19, 20]. The description is valid as long as each physical particle of the system is identical to but distinguishable from the others and as long as the interaction among the particles can be taken to be negligible. Based on the Boltzmann distribution, the probability ($P_i$) that a particle can be found in the $i^{th}$ substate is inversely proportional to the exponential function of the substate energy ($E_i$). A well-known example is the Maxwell–Boltzmann distribution, which describes the velocity distribution of ideal gas molecules [19-21].

The Boltzmann distribution is based on entropy maximization and has been employed in a number of fields. For example, some economists and physicists have introduced entropy concepts into the field of economics and have discussed the distribution of economic systems and their evolution [22-34]. Similarly, econophysicists have employed a stochastic process in describing the dynamics of individual wealth or income and in deriving their probability distributions [33, 35-38].

In this paper, entropy maximization is brought to international emissions trading via the Boltzmann distribution, which provides guidelines for allocating emissions permits among multiple countries. Here, the concept of a physical system is replaced by the concept of an emissions trading system that consists of all participating countries. The concept of the physical particle is replaced by that of the unit emissions permit. The concept of the physical substates is replaced by that of individuals of the participating countries. Assuming that all individuals in a country *i* contribute equally to the total $CO_2$ emissions of that country *i,* the potential energy $E_i$ of a physical substate *i* is replaced by the "allocation potential energy per capita" ($E_i$) of the country *i.* With this replacement, the $E_i$ becomes an *intensive* variable (as apposed to an *extensive* variable) for the country *i* [19]. Thus the probability that a unit emissions permit is allocated to a country *i* is proportional to its total population and is inversely proportional to the exponential function of the allocation potential energy per capita $E_i$. This allocation potential energy per capita may be given in complicated forms and can include various political and economic parameters. One simplest form can be given in a way such that it is negatively proportional to the actual $CO_2$ emissions per capita of a country $i$. Details on the probability distribution function are summarized in Table 1. Table 1 also includes a diagram of the permit allocation using the Boltzmann distribution.

**Table 1**
The Boltzmann distribution for permit allocation

| Boltzmann distribution | Description |
|---|---|
| In physical sciences | $$P_i \propto e^{-\beta E_i}$$ Where, <br> $P_i$ = probability that a particle stays in substate $i$ <br> $e$ = constant of the exponential function ≈ 2.71828 <br> $\beta$ = 1/kT (k = Boltzmann constant, T = absolute temperature) <br> $E_i$ = energy of substate $i$ |
| Potential application In permit allocation | $$P_i \propto C_i e^{-\beta E_i}$$ Where, <br> $P_i$ = probability that emissions permits are allocated to a country $i$ <br> $e$ = constant of the exponential function ≈ 2.71828 <br> $\beta$ = constant ( $\geq$ 0) <br> $E_i$ = allocation potential energy per capita of a country $i$ <br> $C_i$ = total population of a country $i$ |
| Permit allocation using the Boltzmann distribution* | 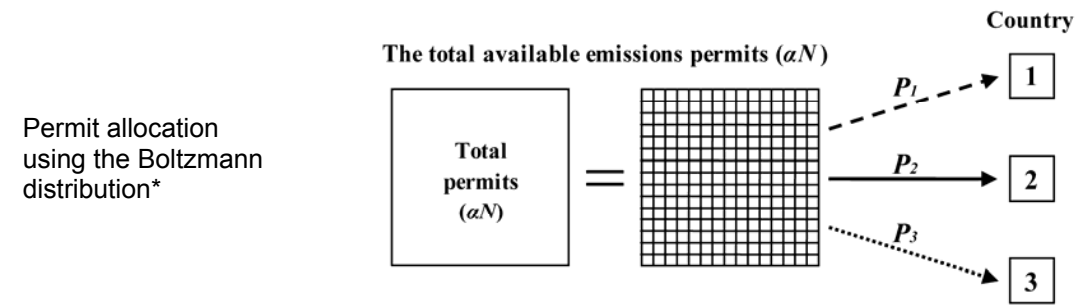 |

\* The total available emissions permits (*αN*) are split into *N* pieces of unit carbon credit *α,* and then the emissions permits are allocated to country *i* (*i* = 1, 2, and 3) based on the probability distribution ($P_i$) from the Boltzmann distribution. Note that the number (*N*) of unit emissions permits can always be made large enough for the Boltzmann statistics by making the unit emissions permit (*α*) smaller.

Note that the Boltzmann distribution in the physical sciences describes a physical system in thermodynamic equilibrium representing the *maximum entropy* at a given total internal energy. In other words, the Boltzmann distribution describes the *most probable* distribution of a physical system. However, the allocation of permits in emissions trading does not have to follow the maximum entropy principle. Indeed, existing allocation methods such as auctioning and grandfathering are unrelated to entropy maximization. What we are attempting to do here is to develop an allocation method that follows the maximum entropy principle using the Boltzmann distribution. In contrast to the existing allocation methods, the new allocation method based on the Boltzmann distribution provides a simple and natural rule for allocating emissions permits among multiple countries.

## 3. The Boltzmann distribution for permit allocation

Suppose that *n* countries are participating in the permit allocation for international emissions trading. Consider that the total available emissions permits ($\alpha N$) are allocated to the countries and that a country *i* has the allocation potential energy per capita of $E_i$ and a population $C_i$ (where *i = 1, 2, 3, . . . , n;* $\alpha$ is the unit emissions permit; and *N* is the total number of unit emissions permits). The number of unit emissions permits that are allocated to the $j^{th}$ individual in a country *i* is $N_i^j$.

We apply two constraints during the permit allocation (Eqs. (1) and (2)).

$$\sum_{i=1}^{n}\sum_{j=1}^{C_i} N_i^j = N \text{ (constant)} \qquad (1)$$

$$\sum_{i=1}^{n}\sum_{j=1}^{C_i} \left(N_i^j E_i\right) = E \text{ (constant)} \qquad (2)$$

The first constraint implies that the total number of available unit emissions permits ($N$) is conserved, that is, that no emissions permits are being taken away or added during allocation. The second constraint corresponds to total energy conservation in the physical sciences. The value $E$ (total global allocation energy) determines the overall distribution of the emissions permits over countries, and therefore, in practice, it needs to be determined based on the agreement among the participating countries. Once determined, it remains constant during the permit allocation. Note that the total allocation energy $E$ is related to the 'temperature' of the system, and therefore to the $\beta$ value in the Boltzmann distribution as described below.

Given the two constraints, the most probable permit allocation following the maximum entropy principle is provided from the Boltzmann distribution, which follows the simple formula [20]

$$N_i^j = Ae^{-\beta E_i}, \qquad (3)$$

where $A$ is a constant and $A$ can be found using the first constraint in Eq. (1). Note that $N_i^j$ is dependent only on $E_i$.

$$\sum_{i=1}^{n}\sum_{j=1}^{C_i} N_i^j = \sum_{i=1}^{n}\sum_{j=1}^{C_i} Ae^{-\beta E_i} = \sum_{i=1}^{n} AC_i e^{-\beta E_i} = A\sum_{i=1}^{n} C_i e^{-\beta E_i} = N \qquad (4)$$

Rearranging,

$$A = \frac{N}{\sum_{i=1}^{n} C_i e^{-\beta E_i}}. \tag{5}$$

Therefore, the probability that a unit emissions permit is allocated to a country $i$ can be expressed in Eq. (6):

$$P_i = \frac{\sum_{j=1}^{C_i} N_i^j}{N} = \frac{C_i N_i^j}{N} = \frac{C_i e^{-\beta E_i}}{\sum_{i=1}^{n} C_i e^{-\beta E_i}}, \quad for\ i = 1, 2, \cdots, n. \tag{6}$$

Then the amount of emissions permits that are allocated to a country $i$ is

$$\alpha N \times P_i = \alpha N \times \frac{C_i e^{-\beta E_i}}{\sum_{i=1}^{n} C_i e^{-\beta E_i}}, \quad for\ i = 1, 2, \cdots, n. \tag{7}$$

In a physical system, the $\beta$ value in the Boltzmann distribution (Eqs. (3-7)) is defined as $1/k\mathrm{T}$ (where $k$ is the Boltzmann constant and T is the temperature of the system). Temperature is directly related to the total energy of the system (the second restraint value ($E$) in Eq. (2)). Therefore, the $\beta$ value is directly related to the overall distribution of the allocated permits and, therefore, it must be determined based on the agreement among the participating countries. In order to provide a guideline for determining the $\beta$ value, we must first understand the influence of the $\beta$ value on permit allocation, which is described below.

## 4. The *β* value in the Boltzmann distribution

The next step is to develop a guideline for determining the *β* value in the Boltzmann distribution. It is assumed that the allocation potential energy per capita ($E_i$) of a country *i* is negatively proportional to the $CO_2$ emissions per capita of a country *i*.

With a *β* value in the positive regime (*β* > 0, which corresponds to the positive temperatures of a physical system), emissions permits tend to be allocated to the countries that have relatively lower allocation potential energy. In this situation, the country with the greatest $CO_2$ emissions per capita (and, therefore, the highest demand for emissions permits) has the lowest allocation potential energy; therefore, it has the largest probability of receiving emissions permits.

To see the effects of the *β* value in permit allocation, we first consider two extreme situations in which *β* values approach the limiting values 0 or ∞. If *β* approaches 0, then the Boltzmann probability in Eq. (6) involves only the populations of the participating countries ($P_i = C_i \big/ \sum_{i=1}^{n} C_i$). In this case, all individuals in the participating countries receive an equal amount of emissions permits, and these are distributed in proportion to the total population of a country *i*. Note that this situation corresponds to one of the suggested fairness notions for the distribution of emissions permits, Egalitarian (in international climate policy, Egalitarian means that "all people have an equal right to pollute or to be protected from pollution") [39, 40]. When *β* approaches 0, the countries with lower $CO_2$ emissions per capita (i.e., higher allocation potential energy per capita) and a large population are most likely satisfied because they can meet their demand relatively easily. On the other hand, this situation is unfavorable to countries with greater $CO_2$ emissions per capita (i.e., lower allocation potential energy per capita). Therefore, the countries with high $CO_2$ emissions per capita will prefer increased $\beta$ values.

If $\beta$ increases and approaches ∞, then the Boltzmann probability acquires non-zero values only for the few countries with the greatest $CO_2$ emissions per capita (lowest allocation potential energy per capita), and becomes 0 for all other countries. In this situation the countries with the greatest $CO_2$ emissions per capita share all the available emissions permits and therefore derive the greatest benefit. But all other countries are left unsatisfied because they obtain only a few permits. Therefore, the other countries will prefer relatively smaller $\beta$ values.

Based on the above situations, we can conclude that the countries with relatively low $CO_2$ emissions per capita and large populations prefer smaller $\beta$ values, and that those with relatively high $CO_2$ emissions per capita prefer larger $\beta$ values. Therefore, agreement on a proper $\beta$ value might not be easily reached, because no single value satisfies all countries. In practice, the $\beta$-value determination will involve a competition between the group of countries having relatively large populations and the group of countries having relatively high $CO_2$ emissions per capita.

In this paper, we do not attempt to derive the single most suitable $\beta$ value for permit allocation. Instead, we suggest a useful reference $\beta$ value using the least square ($y$) calculation between the allocated permits and demand, as in Eq. (8).

$$y = \sum_{i=1}^{n} (permits_i - demand_i)^2 \tag{8}$$

When the least square value ($y$) has its minimum at a $\beta$ value, it can be considered a useful reference $\beta$ value.

## 5. Empirical Data Analysis

To demonstrate permit allocation using the Boltzmann distribution, we selected eight countries. Table 2 shows the $CO_2$ emissions for the eight countries in 2007 and 2008, the $CO_2$ emissions per capita in 2008, and the total population in 2008, the latest years for which annual $CO_2$ emissions data are available, when the following empirical data analysis was carried out. It is assumed that the global (in this case, 8-country) target of $CO_2$ emissions in 2008 is a 3% reduction of the total $CO_2$ emissions in 2007. Thus the total amount of allowed emissions permits for 2008 is 17,084,135 (1000 metric tons), which is then allocated to the eight countries (Canada, China, Germany, Italy, Japan, Russia, the U.K., and the U.S.) using the Boltzmann distribution.

**Table 2**
The carbon dioxide emissions in 2007 and 2008, total population in 2008, and carbon dioxide emissions per capita in 2008.

| Country | 2007 Emissions* (1000 metric tons) | 2008 Emissions* (1000 metric tons) | 2008 Population** | 2008 Emissions per capita (metric tons) |
|---|---|---|---|---|
| Canada | 544,172 | 544,091 | 33,311,400 | 16.33 |
| China | 6,791,805 | 7,031,916 | 1,324,655,000 | 5.31 |
| Germany | 787,235 | 786,660 | 82,110,097 | 9.58 |
| Italy | 459,376 | 445,119 | 59,832,179 | 7.44 |
| Japan | 1,251,188 | 1,208,163 | 127,704,040 | 9.46 |
| Russia | 1,667,576 | 1,708,653 | 141,950,000 | 12.04 |
| UK | 529,621 | 522,856 | 61,393,521 | 8.52 |
| US | 5,581,537 | 5,461,014 | 304,375,000 | 17.94 |
| Total | 17,612,510*** | 17,708,472 | 2,135,331,237 | 86.62 |

\* Source: Millennium Development Goals Database in United Nations Statistics Division
The most recent data (2008) are used in this paper.
See http://data.un.org/Data.aspx?q=co2+emissions&d=MDG&f=seriesRowID%3a749
\*\* Source: World Development Indicators in the World Bank
See
http://data.un.org/Data.aspx?q=population+datamart%5bWDI%5d&d=WDI&f=Indicator_Code%3aSP.POP.TOTL
\*\*\* 97 percent of 2007 emissions (17,084,135 (1000 metric tons)) was set to be the global target and the value of the total available emissions permits for 2008 in the empirical data analysis.

In the following empirical data analysis, the allocation potential energy per capita ($E_i$) of a country $i$ is defined as the negative value of $CO_2$ emissions per capita of a country $i$ in 2008. Using this simple definition, the emissions permits are more likely to be allocated to the countries with greater $CO_2$ emissions per capita.

As described earlier, the permit allocation based on the Boltzmann distribution is largely affected by the $β$ value in Eq. (6). Therefore, we first consider a reasonable range of $β$ values for the permit allocation for each country. Fig. 1 shows the permit allocation using the Boltzmann distribution as a function of the $\beta$ value. From this data, the $\beta$-value range preferred by each country can be determined. As shown in Fig. 1, the U.S. meets its demand ($CO_2$ emissions in 2008) at the $\beta$ value of 0.0953. Therefore, the U.S. prefers values larger than 0.0953, and Canada prefers $\beta$ values between 0.1063 and 0.7426. China prefers $\beta$ values smaller than 0.0978, and Italy prefers $\beta$ values smaller than 0.0507. But the four remaining countries (Germany, Japan, Russia, and the U.K.) cannot obtain sufficient emissions permits to meet their actual emissions. This is because the total allowed emissions permits in 2008 (97% of 2007 emissions) is less than the total $CO_2$ emissions in 2008. Note that, as shown in Fig. 1, the permit allocation is driven mainly by the competition between two countries: China, which produces the greatest $CO_2$ emissions overall (but has the smallest $CO_2$ emissions per capita due to its large population), and the U.S., which has the greatest $CO_2$ emissions per capita.

This observation revisits a challenging question in the reduction of $CO_2$ emissions. Which country should be most responsible for future reduction: the one that produces the greatest $CO_2$ emissions overall (China), or the one that produces the greatest $CO_2$ emissions per capita (the U.S.)? In the permit allocation using the Boltzmann distribution, the responsibility can be shifted between China and the U.S. by changing the $\beta$ value, as shown in Fig. 1. Below a $\beta$ value of 0.1164, China receives more permits than the U.S., thus leaving more responsibility to the U.S. Above that $\beta$ value, the U.S. receives more permits than China, thus leaving more responsibility to China. This result shows the flexibility of allocating permits using the Boltzmann distribution and the importance of determining a proper *β* value.

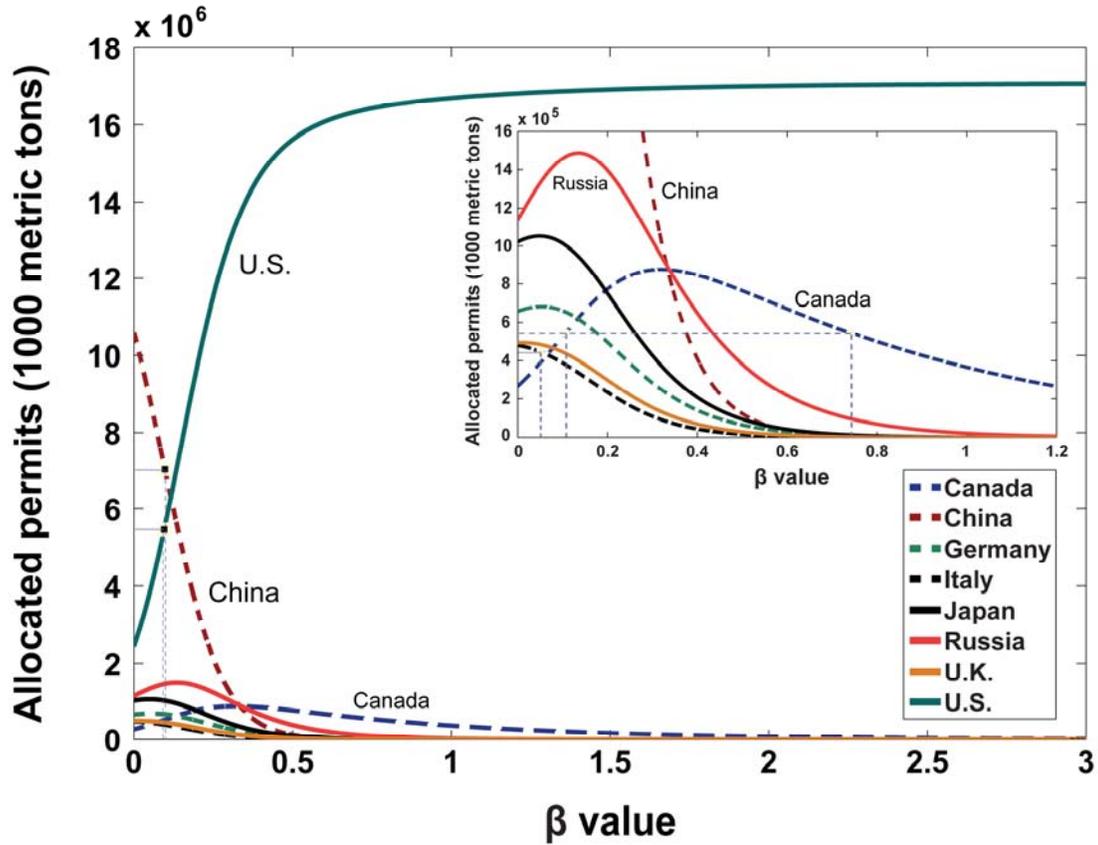

**Fig. 1**. Permit allocation among eight countries using the Boltzmann distribution as a function of the *β* value. The emissions permits allocated to the U.S. are smallest when the *β* value is 0. The amount rapidly increases up to the *β* value of ~0.3, and beyond ~0.3 it increases rather slowly and approaches the total amount of available emissions permits. The amount of emissions permits allocated to China is the largest when the *β* value is 0. The amount rapidly decreases up to the *β* value of ~0.3 and approaches 0. The amount of emissions permits allocated to Italy is the largest when the *β* value is 0, and the amount monotonously decreases up to the *β* value of ~0.3. For all other countries (Canada, Germany, Japan, Russia, and the U.K.), the amount first increases and then gradually decreases to 0 in the given *β*-value range. The gray dotted line for China, the U.S., Canada, and Italy shows the *β* value at which the allocated permits meets their actual $CO_2$ emissions in 2008.

Fig. 2 shows the least square ($y$) calculation (Eq. (8)) between allocated permits and demand (actual $CO_2$ emissions in 2008) as a function of the $\beta$ value. In this empirical data analysis, the least square value ($y$) has its minimum at a $\beta$ value of 0.0966. The results of emissions permit allocation with the reference $\beta$ value are summarized in Table 3. The probability values for permit allocation are distributed between 0.02 and 0.41 for the eight countries, and the number of emissions permits allocated to the countries ranges from 392,870 to 7,079,729 (in 1000 metric tons).

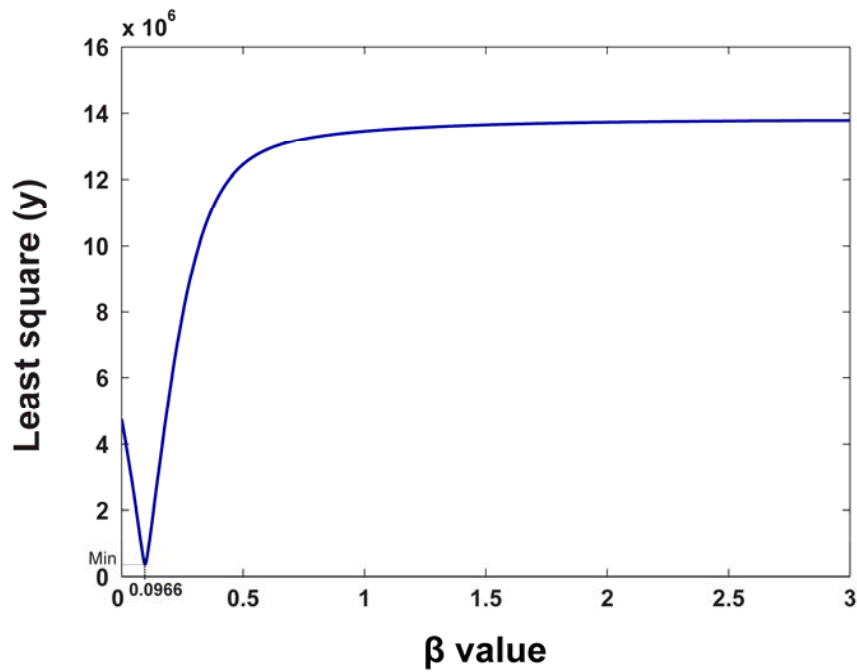

**Fig. 2.** Least square (y) calculation between allocated permits and demand (actual $CO_2$ emissions in 2008) as a function of the $\beta$ value. The least square value ($y$) has its minimum at the $\beta$ value of 0.0966, at which the difference between the allocated permits and demand becomes the smallest for the eight participating countries. This $\beta$ value provides a reference point for the permit allocation.

Fig. 3 shows permit allocation using the Boltzmann distribution with the reference $\beta$ value ($\beta$ = 0.0966), and the allocated permits are compared with their actual $CO_2$ emissions in 2008. As listed in Table 3 and shown in Fig. 3, the permits allocated to two countries (China and the U.S.) exceed their demands. For the remaining six countries (Canada, Germany, Italy, Japan, Russia, and the U.K.), the situation is the reverse. Therefore, after this initial permit allocation, emissions trading can occur between permit sellers (China and the U.S.) and permit buyers (Canada, Germany, Italy, Japan, Russia, and the U.K.). It is also possible that extra emissions permits can be considered and supplied through the developing of technology, the planting of trees, and the reducing of $CO_2$ emissions by carbon tax during emissions trading [14, 41-44]. Note that this empirical data analysis shows permit allocation among eight countries. If more countries are added, then permit sellers and permit buyers can be changed.

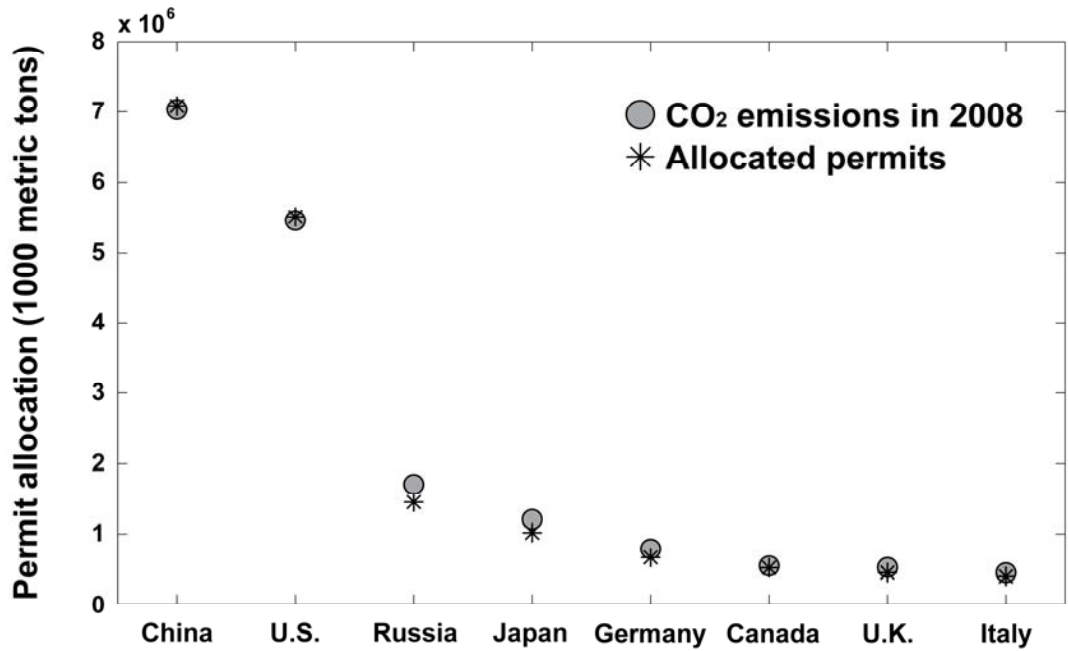

**Fig. 3.** Permit allocation using the Boltzmann distribution with the least square reference $β$ value ($β = 0.0966$). It shows the $CO_2$ emissions in 2008 of the participating countries and the emissions permits allocated by the Boltzmann distribution. After the permit allocation, emissions permits can be traded between countries.

**Table 3**
The emissions permits of eight countries using the Boltzmann distribution ($β = 0.0966$).

| Country | $CO_2$ emissions in 2008 (1000 metric tons) | Boltzmann probability | Allocated permits (1000 metric tons) | Difference* |
|---|---|---|---|---|
| Canada | 544,091 | 0.03 | 516,460 | -27,631 |
| China | 7,031,916 | 0.41 | 7,079,729 | 47,813 |
| Germany | 786,660 | 0.04 | 663,034 | -123,626 |
| Italy | 445,119 | 0.02 | 392,870 | -52,249 |
| Japan | 1,208,163 | 0.06 | 1,019,327 | -188,836 |
| Russia | 1,708,653 | 0.09 | 1,453,215 | -255,438 |
| U.K. | 522,856 | 0.03 | 447,322 | -75,534 |
| U.S. | 5,461,014 | 0.32 | 5,512,179 | 51,165 |
| Total | 17,708,472 | 1.00 | 17,084,135 | -624,337 |

* Difference = (allocated permits – actual carbon dioxide emissions in 2008).

## 6. Discussions and Conclusions

The Boltzmann distribution, originating in the physical sciences, is based on entropy maximization and thus provides the *most probable* distribution of a physical system at equilibrium. Banerjee and Yakovenko (2010) observed that social and economic inequality is ubiquitous in the real world, and showed that the common theme of three specific cases (e.g., the distribution of money, income, and global energy consumption) is entropy maximization for the partitioning of a limited resource among multiple agents [45].

In this paper, we applied entropy maximization to the problem of permit allocation in emissions trading. When brought to initial permit allocation, the Boltzmann distribution provides the *most probable* allocation among multiple countries. The concept of *most probable* in the physical sciences may be translated into *fair* in permit allocation, as the distribution provides a natural and undistorted allocation among participating countries. We showed that the Boltzmann distribution includes one of the fairness notions, Egalitarian, when the β value approaches 0.

Throughout this paper, we developed a mathematical description of permit allocation using the Boltzmann distribution. Then, through an empirical data analysis, we demonstrated how initial permit allocation can be performed over eight countries using the Boltzmann distribution. If emissions permits are allocated to the participating countries free of charge, the Boltzmann allocation is more like the grandfathering method, in which emissions permits are distributed freely based on the historical output of each country. If emissions permits are allocated to the countries at their expense (for example, proportional to the amount of allocated permits), then the Boltzmann allocation is similar to auctioning. However, in the Boltzmann allocation, the cartel of bidders and speculative behaviors observed in auctioning become more limited. Further studies are needed to develop a proper price model for the Boltzmann allocation. In addition, more work is needed to find a sophisticated definition of the allocation potential energy per capita that reflects a variety of political and economical parameters.

Note that the permit allocation using the Boltzmann distribution described here is a simple yet versatile, flexible method. In other words, it can be modified and adjusted for a variety of different purposes in addition to permit allocation. Two fundamental facts, limited resources and unlimited human wants, provide a foundation for the field of economics [46]. Where resources are limited or scarce and all human wants are virtually unlimited or insatiable, the Boltzmann distribution might be applicable. Such situations include various economic and environmental problems, such as the tradable permits approach to protect the commons (e.g., allowances of water and air pollution control, rights in water supply management, and quotas in fisheries management). In addition, the Boltzmann distribution might be applicable to fair division, also known as the cake-cutting problem.

There are numerous methods for obtaining a fair division, or dividing a resource in such a way that all recipients believe they have received a fair amount (e.g., proportional fair division, envy-free division, exact division, Pareto optimal division, and equitable division) [47-53]. In the simplest two-player case, the well-known procedure of "the divide and choose method" leads to a proportional and envy-free division [54, 55]. However, this case may not consider other essential factors such as the weight or the daily required caloric intake of players. Furthermore, fair division with three or more players is considerably more complex than the two-player case [47, 48, 51, 56-58]. For example, let's assume a cake is to be divided and distributed among three players: two adults, one of whom weighs 100 kgs and the other 55 kgs, and a child who weighs 20 kgs. If this cake is cut into three identical pieces, is it a fair division for the players?

The division of a single homogeneous good (for example, a cake) among multiple players is easily accomplished using the Boltzmann distribution as described in this paper. In this case, an individual player's weight or other essential factors (for example, daily required caloric intake) can be considered his or her allocation potential energy. For example, the allocation potential energy can be defined as the daily required caloric intake per unit weight. The cake can then be distributed to the players, taking into consideration their total weight and following the Boltzmann distribution probability.

In summary, the Boltzmann distribution can be applied not only to permit allocation in emissions trading but also to other economic (e.g., fair division) and environmental (e.g., air and water pollution control, water and food supply management, and fisheries management) problems by replacing the allocation potential energy with the proper economic concepts. In this sense, the Boltzmann distribution based on entropy maximization may become a useful method for a variety of economic and environmental applications.

## Acknowledgments

We thank Timothy Mount and Kieran P. Donaghy (Cornell University) for thoughtful comments.